\documentclass[pdflatex,sn-mathphys-num]{sn-jnl}

\usepackage{graphicx}%
\usepackage{multirow}%
\usepackage{amsmath,amssymb,amsfonts}%
\usepackage{amsthm}%
\usepackage{mathrsfs}%
\usepackage[title]{appendix}%
\usepackage{xcolor}%
\usepackage{textcomp}%
\usepackage{manyfoot}%
\usepackage{booktabs}%
\usepackage{url}
\usepackage{algorithm}%
\usepackage{algorithmicx}%
\usepackage{algpseudocode}%
\usepackage{listings}%
\usepackage{xcolor}

\usepackage{afterpage,lscape}

\usepackage{array}
\newcolumntype{L}[1]{>{\raggedright\arraybackslash}p{#1}}

\usepackage{comment} 

\theoremstyle{thmstyleone}%
\theoremstyle{thmstyletwo}%

\theoremstyle{thmstylethree}%

\begin{document}


\title[CECAM Biochar]{What makes a useful molecular model of biochar? A community roadmap}


\author[1]{\fnm{Valentina} \sur{Sierra-Jimenez}}\email{v.sierrajimenez@wsu.edu}

\author[2]{\fnm{Jonathan P.} \sur{Mathews}}\email{jmathews@psu.edu}

\author[3]{\fnm{Luca} \sur{Bellucci}}\email{luca.bellucci@nano.cnr.it}

\author[4]{\fnm{Edo} \sur{Boek}}\email{e.boek@qmul.ac.uk}

\author[5]{\fnm{Carla} \sur{de Tomas}}\email{carla.de\_tomas@kcl.ac.uk}

\author[1]{\fnm{Manuel} \sur{Garcia-Perez}}\email{mgarcia-perez@wsu.edu}

\author[6]{\fnm{Stef} \sur{Ghysels}}\email{stef.ghysels@ugent.be}

\author[7]{\fnm{Paola} \sur{Giudicianni}}\email{paola.giudicianni@stems.cnr.it}

\author[7]{\fnm{Corinna Maria} \sur{Grottola}}\email{corinnamaria.grottola@stems.cnr.it}

\author[8]{\fnm{Kelly Anne} \sur{Hawboldt}}\email{khawboldt@mun.ca}

\author[9]{\fnm{Robert L.} \sur{Johnson}}\email{robertlj@hawaii.edu}

\author[10]{\fnm{Fenna B.E.} \sur{Kolff}}\email{f.b.e.kolff@tudelft.nl}

\author[11]{\fnm{Jean-Marc} \sur{Leyssale}}\email{jean-marc.leyssale@u-bordeaux.fr}

\author[12]{\fnm{Diego} \sur{Liberati}}\email{diego.liberati@cnr.it}

\author[13]{\fnm{Francisco J.} \sur{Martin-Martinez}}\email{francisco.martin-martinez@kcl.ac.uk}

\author[14]{\fnm{Jacob W.} \sur{Martin}}\email{jacob.w.martin@curtin.edu.au}

\author[15]{\fnm{Ond\v{r}ej} \sur{Ma\v{s}ek}}\email{ondrej.masek@ed.ac.uk}

\author[1,16]{\fnm{Mohammad} \sur{Mezbah Ul Hoque}}\email{m.hoque1@wsu.edu}

\author[15,17]{\fnm{Audrey} \sur{Ngambia}}\email{a.l.noumbissi-ngambia@ed.ac.uk}

\author[11]{ \fnm{Ama\"el} \sur{Obliger}}\email{amael.obliger@u-bordeaux.fr}

\author[18]{ \fnm{Frederik} \sur{Ossler}}\email{frederik.ossler@fysik.lu.se}

\author[19]{\fnm{Muhammad} \sur{Riaz}}\email{muhammad.riaz@gcuf.edu.pk}

\author[17]{\fnm{John M.} \sur{Tobin}}\email{j.tobin@ed.ac.uk}

\author[20]{\fnm{Xiaolei} \sur{Zhang}}\email{xiaolei.zhang@strath.ac.uk}

\author*[17]{\fnm{Valentina} \sur{Erastova}}\email{valentina.erastova@ed.ac.uk}


\affil[1]{\orgdiv{Department of Biological Systems Engineering}, \orgname{Washington State University}, \orgaddress{\street{1935 E. Grimes Way}, \city{Pullman}, \postcode{99164-6120}, \state{WA}, \country{USA}}}

\affil[2]{\orgdiv{John and Willie Leone Family Department of Energy and Mineral Engineering, and the Earth and Mineral Sciences Energy Institute}, \orgname{The Pennsylvania State University}, \orgaddress{\street{58 Pollock Street}, \city{University Park}, \postcode{16802}, \state{PA}, \country{USA}}}

\affil[3]{\orgdiv{Istituto Nanoscienze-CNR}, \orgname{NEST-SNS}, \orgaddress{\street{Piazza San Silvestro 12}, \city{Pisa}, \postcode{56127}, \country{Italy}}}

\affil[4]{\orgdiv{School of Engineering and Materials Science}, \orgname{Queen Mary University of London}, \orgaddress{\street{Mile End Road}, \city{London}, \postcode{E1 4NS}, \country{UK}}}

\affil[5]{\orgdiv{Department of Physics}, \orgname{King's College London}, \orgaddress{\street{Strand}, \city{London}, \postcode{WC2R 2LS}, \country{UK}}}

\affil[6]{\orgdiv{Department of Green Chemistry and Technology, Faculty of Bioscience Engineering}, \orgname{Ghent University}, \orgaddress{\street{Coupure Links 653}, \city{Ghent}, \postcode{9000}, \country{Belgium}}}

\affil[7]{\orgdiv{Institute of Science and Technologies for Sustainable Energy and Mobility (STEMS)}, \orgname{National Research Council (CNR)}, \orgaddress{\street{Via Guglielmo Marconi, 4}, \city{Naples}, \postcode{80125}, \country{Italy}}}

\affil[8]{\orgdiv{Department of Process Engineering, Faculty of Engineering and Applied Science}, \orgname{Memorial University of Newfoundland}, \orgaddress{\street{230 Elizabeth Ave}, \city{St. John's}, \postcode{A1B 3X5}, \state{NL}, \country{Canada}}}

\affil[9]{\orgdiv{Hawaii Natural Energy Institute}, \orgname{University of Hawaii at M\=anoa}, \orgaddress{\street{1680 East-West Road}, \city{Honolulu}, \postcode{96822}, \state{HI}, \country{USA}}}

\affil[10]{\orgdiv{Department of Process \& Energy}, \orgname{Delft University of Technology}, \orgaddress{\street{Leeghwaterstraat 39}, \city{Delft}, \postcode{2628CB}, \country{The Netherlands}}}

\affil[11]{\orgdiv{Institut des Sciences Mol\'eculaires}, \orgname{Universit\'e de Bordeaux, CNRS, Bordeaux INP}, \orgaddress{\street{351 Cours de la Lib\'eration}, \city{Talence}, \postcode{33405}, \country{France}}}

\affil[12]{\orgdiv{National Research Council of Italy (CNR), Institute of Electronics, Information and Communication Engineering}, \orgname{Politecnico di Milano}, \orgaddress{\street{Piazza Leonardo da Vinci 32}, \city{Milano}, \postcode{20133}, \country{Italy}}}

\affil[13]{\orgdiv{Department of Chemistry}, \orgname{King's College London}, \orgaddress{\street{Strand}, \city{London}, \postcode{WC2R 2LS}, \country{UK}}}

\affil[14]{\orgdiv{Discipline of Physics and Astronomy}, \orgname{Curtin University}, \orgaddress{\street{Kent Street, Bentley}, \city{Perth}, \postcode{6102}, \state{Western Australia}, \country{Australia}}}

\affil[15]{\orgdiv{School of GeoSciences}, \orgname{University of Edinburgh}, \orgaddress{\street{Alexander Crum Brown Road}, \city{Edinburgh}, \postcode{EH9 3FF}, \country{UK}}}

\affil[16]{\orgdiv{Composite Materials and Engineering Center}, \orgname{Washington State University}, \orgaddress{\street{ 2001 Grimes Way}, \city{Pullman}, \postcode{99164-6120}, \state{WA}, \country{USA}}}

\affil[17]{\orgdiv{School of Chemistry}, \orgname{University of Edinburgh}, \orgaddress{\street{Joseph Black Building, David Brewster Road}, \city{Edinburgh}, \postcode{EH9 3FJ}, \country{UK}}}

\affil[18]{\orgdiv{Department of Physics}, \orgname{Lund University}, \orgaddress{\city{Lund},  \postcode{22100}, \country{Sweden}}}

\affil[19]{\orgdiv{Department of Environmental Sciences}, \orgname{Government College University Faisalabad}, \orgaddress{\street{Allama Iqbal Road}, \city{Faisalabad}, \postcode{38000}, \country{Pakistan}}}

\affil[20]{\orgdiv{Department of Chemical and Process Engineering}, \orgname{University of Strathclyde}, \orgaddress{\street{75 Montrose Street}, \city{Glasgow}, \postcode{G1 1XJ}, \country{UK}}}













\abstract{ 
Biochars are disordered carbonaceous materials produced by biomass pyrolysis with applications spanning soil amendment, water remediation, long-term carbon storage, adsorption, and functional materials. 
Although they share important structural and chemical features with other disordered carbons, including coal, kerogen, and activated carbons, the questions posed to biochar models are distinct. 
A model built to predict soil persistence must capture reactive surface chemistry, water accessibility, and biological exposure; a model built to screen gas separation or aqueous remediation must instead reproduce pore architecture, solvation, competitive ions, and adsorption energetics.
No single model can achieve all of this equally well. Model usefulness must therefore be defined relative to a specific question and validated against independent experimental observables.

This community roadmap, arising from a CECAM workshop on advancing biochar molecular models, critically maps current molecular and atomistic approaches, spanning experimentally guided top-down reconstruction, mimetic bottom-up simulation, and hybrid methods that combine mechanistic construction with experimental constraints.

We argue that the field's first generation of models has been more successful than is often acknowledged, provided they are built at a sufficiently large length scale and with explicit control over microporosity and bulk chemical functionality.
Structural and equilibrium interfacial properties are increasingly tractable with classical, non-reactive force fields, whereas dynamic and reactive behaviours remain poorly described and will require selective use of reactive methods within multiscale workflows rather than a single universal model.
A parallel and largely unaddressed gap concerns the mineral and ash components of biochar -- silicates, carbonates, phosphates, and metal oxides that vary widely with feedstock and pyrolysis conditions -- and the changes the whole material undergoes during ageing in soil. 
Models of the organic carbon framework describe only part of the material; incorporating mineral phases and their interfaces with carbon is essential for soil chemistry, nutrient release, and long-term carbon persistence predictions.

We identify \textit{seven open questions} that current models cannot yet answer reliably: 
    carbon persistence, 
    resistance to mechanical and biological breakdown, 
    heteroatom-controlled adsorption and catalytic selectivity,
    carbon--mineral interfacial chemistry, 
    molecular-sieve behaviour, 
    ageing and surface oxidation
    and the surface--bulk boundary. 
We propose \textit{five community priorities}: 
    force field benchmarking and development, 
    open model and data repositories, 
    shared classification and metadata standards,
    ensemble validation, 
    and training in reproducible modelling practice.
Across these priorities, sustained interaction with experimentalists is essential, both to ground models in real observables and to document where models fail as clearly as where they succeed. 
Progress will be accelerated by continuing to adapt transferable methods from coal, kerogen, clay--organic matter, and other disordered-carbon frameworks rather than repeating their trial-and-error development. 
}



\maketitle


\section*{List of Abbreviations}

\begin{description}
\item[BET] Brunauer--Emmett--Teller
\item[BPCA] Benzene Polycarboxylic Acids
\item[CECAM] Centre Europ\'een de Calcul Atomique et Mol\'eculaire
\item[CHARMM] Chemistry at Harvard Macromolecular Mechanics
\item[DFT] Density Functional Theory
\item[EDX] Energy-Dispersive X-ray Spectroscopy
\item[EPR] Electron Paramagnetic Resonance
\item[FT-ICR MS] Fourier Transform Ion Cyclotron Resonance Mass Spectrometry
\item[FTIR] Fourier Transform Infrared Spectroscopy
\item[GCMC] Grand Canonical Monte Carlo
\item[H/C\textsubscript{org}] Atomic ratio of hydrogen to organic carbon
\item[HR-TEM] High-Resolution Transmission Electron Microscopy
\item[ICP-MS] Inductively Coupled Plasma Mass Spectrometry
\item[ICP-OES] Inductively Coupled Plasma Optical Emission Spectrometry
\item[LDI] Laser Desorption Ionisation
\item[MACE] Multi-Atomic Cluster Expansion
\item[MC] Monte Carlo
\item[MD] Molecular Dynamics
\item[micro-CT] Micro-Computed Tomography
\item[ML] Machine Learning
\item[NMR] Nuclear Magnetic Resonance
\item[O/C\textsubscript{org}] Atomic ratio of oxygen to organic carbon
\item[OPLS] Optimized Potentials for Liquid Simulations
\item[PDF] Pair Distribution Function
\item[pH\textsubscript{pzc}] pH at the Point of Zero Charge
\item[R50] Thermal Recalcitrance Index Based on Relative Oxidation Resistance
\item[ReaxFF] Reactive Force Field
\item[SANS] Small-Angle Neutron Scattering
\item[SAXS] Small-Angle X-ray Scattering
\item[SEM] Scanning Electron Microscopy
\item[TEM] Transmission Electron Microscopy
\item[TGA] Thermogravimetric Analysis
\item[TPO] Temperature-Programmed Oxidation
\item[TraPPE] Transferable Potentials for Phase Equilibria
\item[WAXS] Wide-Angle X-ray Scattering
\item[XANES] X-ray Absorption Near-Edge Structure
\item[XPS] X-ray Photoelectron Spectroscopy
\item[XRD] X-ray Diffraction
\end{description}


\section{Biochar is not simply charcoal}\label{sec1}

Biochar occupies overlapping chemical space with charcoal: both are carbon-rich materials produced by thermal transformation of biomass, yet their intended functions differ. 
Charcoal is primarily a fuel, whereas biochar is produced for soil application, carbon storage, or upgrading into functional materials. 
That shift in purpose changes what a model of a biochar material must be able to capture. 
Many examples discussed below are drawn from woody and cellulosic biochars because these lower-mineral systems are where atomistic models and validation datasets are currently most developed, not because they span the full biochar class. 
The broader biochar class also includes herbaceous, manure-derived, chitin-rich, and other biomass-derived chars.

Four application domains drive biochar research.
As \textit{a soil amendment}, biochar improves fertility, water retention, soil aeration, and agricultural sustainability, building on centuries of indigenous practice in the Terra Preta dark earths of the Amazon basin.\cite{marris2006putting}
As \textit{a carbon sink}, biochar offers a route to long-term atmospheric carbon storage, provided it resists biological and physical degradation over decades or centuries. Its stability and soil interactions are also shaped by mineral and ash components inherited from the feedstock and modified during pyrolysis.\cite{lehmann2006bio}
As \textit{a sorbent for water treatment and environmental remediation}, biochar sequesters heavy metals, nutrients, and organic micropollutants from aqueous streams.
Unlike dry-state applications, this wet-environment domain requires managing competitive ion effects, pH-dependent surface charges, and fouling by dissolved organic matter.\cite{ahmad2014biochar}
As \textit{a precursor to advanced materials} -- supercapacitors, catalysts, composites, battery anodes, 
fillers, and adsorbents -- it provides a renewable alternative to fossil-derived carbons and, in some cases, a heteroatom-containing starting point for functional porous materials.\cite{lopez2019conductive, xiao2018insight, bachs2023biomass} 

Biochar is not a pure carbon material. 
Hydrogen remains in the carbonised matrix, with atomic H/C ratios generally decreasing as thermal treatment severity increases and depending on feedstock, heating rate, solid residence time, and, to a lesser extent, pressure.\cite{ghysels2019production}
Heteroatoms such as O, N, and S are also inherently present and may be further enriched in the material by feedstock choice, feedstock pretreatment, pyrolysis conditions, or post-production doping.\cite{haghighi2022perspectives}
Depending on the feedstock, ash content can range from below 1 wt.\% in woody biochars to above 40 wt.\% in grass- and manure-derived biochars.\cite{Lehmann2024}
Mineral phases -- crystalline and amorphous silicates, calcium and magnesium carbonates, potassium salts, iron oxides, or phosphates -- are not inert spectators. 
During thermal treatment they help shape the organic phase itself: biochars with similar H/C and O/C ratios may still carry different functional-group distributions, aromatic connectivities, and surface chemistries, depending on feedstock mineralogy. 
Once formed, mineral phases continue to modify pore geometry, alter surface potential, and provide reactive sites for metal complexation, dissolving at different rates as conditions change or the material ages in soil.\cite{sowers2018spatial, varga2025role}

The properties driving biochar applications span length scales. 
Macropores inherited from plant tissue, such as vessels and cell lumens, can control transport at tens to hundreds of micrometres, whereas adsorption sites, functional groups, nanopores, and carbon--mineral interfaces are nanometre-scale features. 
Molecular models do not represent the entire particle; they provide nanoscale structural and chemical motifs that must ultimately be passed to coarse-grained, particle-scale, or continuum descriptions.

Each application asks fundamentally different questions of the material. For example: 
\textit{How long does biochar persist in soil?} 
\textit{What pore geometry enables gas separation?} 
\textit{How do surface functional groups drive catalytic selectivity?} 
\textit{How do water chemistry and competitive ions control contaminant removal?} 
What a model must do depends entirely on what question it is intended to answer.

Biochar research does not begin from a blank slate. 
Coal, kerogen, soot, activated carbon, and carbon molecular sieve communities have already developed modelling tools for chemically analogous disordered carbons, including force fields, reconstruction workflows, structural descriptors, and validation strategies.\cite{SierraJimenez2023review} 
These communities have shown that both experimentally guided top-down reconstruction and mechanistically driven bottom-up simulation can generate atomistic representations of carbonaceous materials.\cite{FernandezAlos2011, ungerer2015molecular, Leyssale2017, Obliger2023, luo2021virtual} 
Although these methods were not designed for biochar, they transfer most directly to its organic carbon framework, whereas mineral phases, carbon--mineral interfaces, and ageing remain comparatively underdeveloped. 
Because these gaps span pyrolysis chemistry, porous carbon physics, environmental geochemistry, soil science, and molecular simulation, we frame this roadmap as a community synthesis rather than a single-group prescription.


\section{The role of models} \label{sec2}

Any model is a simplified description of a target aspect of reality. 
A poor model makes unreliable predictions. 
A useful model makes reliable, question-specific predictions within a clearly stated domain of applicability, and is honest about where that domain ends.
As Box observed, all models are wrong -- the question is whether they are useful.\cite{box1979robustness}
Reliability, however, is not established by chemical plausibility or visual resemblance alone.
Where possible, models should be assessed against independent experimental observables not used in construction; where direct validation is unavailable, confidence should come from transferability across related questions without re-fitting. Extrapolating beyond that domain is a hypothesis, not a validation.
For descriptors that remain difficult to measure directly, such as active-site density, accessibility classes, carbon--mineral interfaces, or ageing trajectories, models should report the experimental proxies used rather than presenting the descriptor as independently validated.
XPS illustrates the principle directly: atomistic models help deconvolve overlapping C 1s and O 1s signals, while the resulting spectra in turn constrain and challengefunctional-group assignments fed back into the structure.\cite{zarrouk2024experiment}

Molecular and atomistic models describe matter at the nanoscale: individual atoms, bonds, and non-covalent interactions. 
DFT resolves electronic structure and local reaction energetics, whereas MD samples statistical structure, hydration, adsorption, and transport in larger biochar--molecule ensembles.
The nanoscale is not the only relevant scale: continuum models describe field-scale transport, and finite-element methods describe particle-scale mechanical response. 
It is at the nanoscale, however, that the properties distinguishing one biochar from another originate -- surface chemistry, nanopore geometry, carbon-network disorder, heteroatom distribution, and, for many agronomic biochars, mineral inclusions and carbon--mineral interfaces.
Getting these right is the prerequisite for everything above.

Molecular modelling has already made the nanoscale tractable across biochemistry and solid-state physics.
Protein folding, semiconductor band gaps, and binding affinities became reliably modelled quantities not because the models were perfect, but because they were useful within defined limits. 
Biochar is structurally more disordered and chemically more heterogeneous than many canonical systems, which makes model choice -- and honesty about assumptions -- even more consequential for model reliability.
For disordered carbons, usefulness often lies in an ensemble of experimentally consistent structures rather than in a single unique atomistic model. 
Even a simple model earns its place if it narrows the experimental search space or flags where intuition is likely wrong, rather than only if it predicts a final number correctly.

\textit{The scientific question must drive the model.}
A model built to predict carbon longevity in soil must capture oxidative surface chemistry, accessibility, and eventually ageing.
A model built to screen gas adsorbents must accurately reproduce pore geometry, size, distribution, and adsorption energetics.
A model built for aqueous remediation or analytical pre-concentration must additionally represent water structure, competitive ions, pH-dependent charge, and matrix effects.
No single model does everything, and expecting it to is how models lose credibility.


\section{Current approaches to constructing biochar molecular models} \label{sec3}

Biochar modelling began with pen and paper. 
Early models were hand-drawn arrangements of aromatic rings and functional groups -- useful for communicating structural intuition, useless for simulations. 
The transition to quantitative atomistic models, with explicit coordinates and interatomic potentials, marked a genuine advance. 
Yet, it also introduced a new failure mode -- the appearance of rigour without substance.
A molecular simulation of a poorly constructed structural model produces numbers that look like predictions but are not independently reliable.

\begin{landscape}
\begin{figure}[]
      \centering
      \includegraphics[width=1\linewidth]{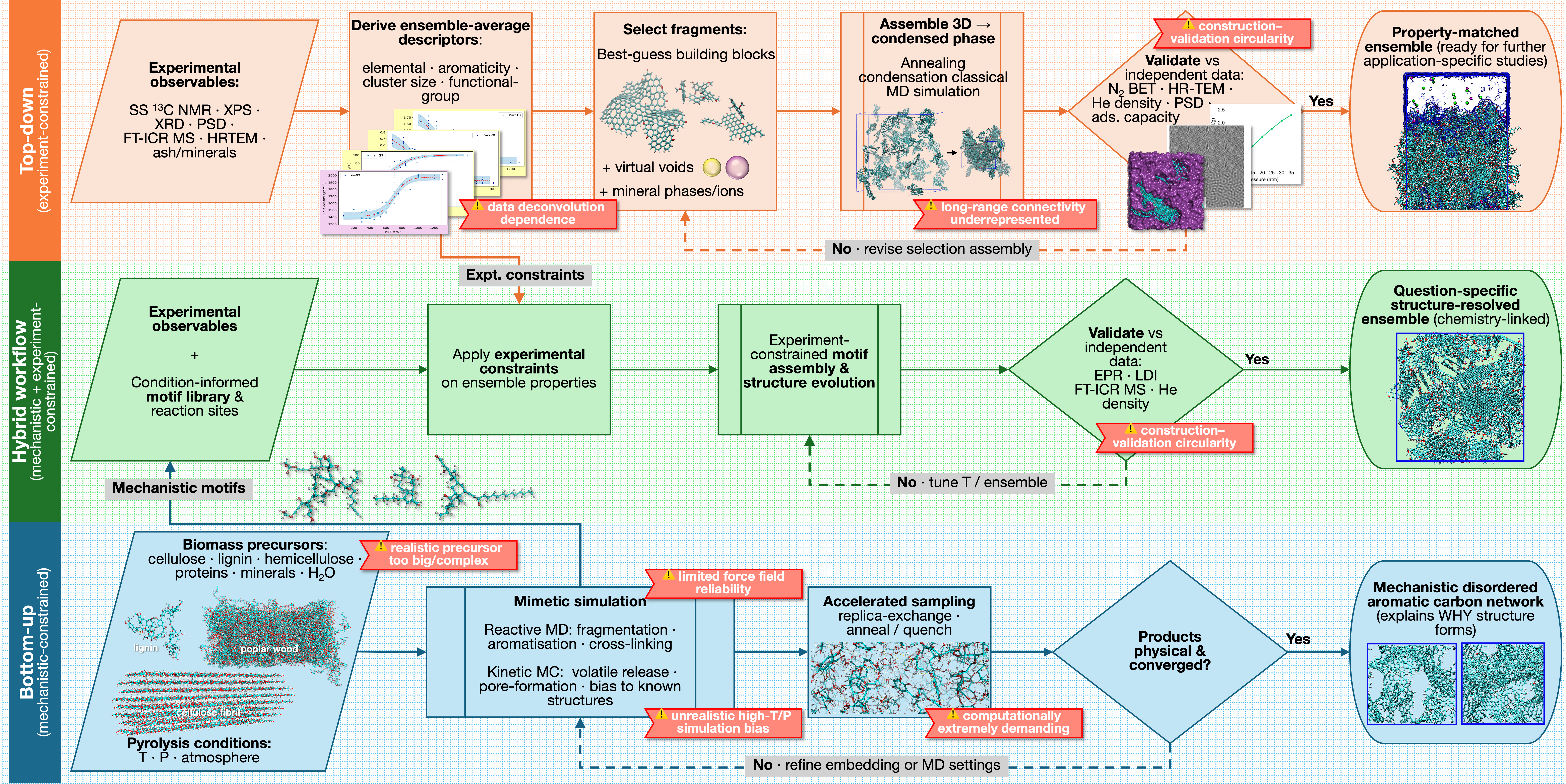}
      \caption{Schematic comparison of experiment-constrained, mechanistic, and hybrid routes to biochar molecular model construction. 
      Top-down workflows use experimental observables to constrain molecular fragments and their assembly. 
      Bottom-up workflows generate structural motifs mechanistically from biomass precursors or pyrolysis products. 
      Hybrid workflows combine these routes by using mechanistic simulations to identify motifs and reactive sites, then embedding them in experimentally constrained models validated against independent observables. Structures shown are taken from authors own work.}
      \label{fig:approaches}
  \end{figure}  
\end{landscape}

Constructing a structural model that avoids this failure mode is the shared aim of current workflows, which draw on two end-member strategies -- top-down and bottom-up -- and, increasingly, on hybrid combinations that sit between them. 
Figure \ref{fig:approaches} maps their inputs, outputs, and overlap. 
Because experiments do not uniquely determine a biochar atomistic structure, the practical target of any of these routes is usually a property-matched representative ensemble rather than a single exact reconstruction.\cite{bellucci2020silico}

Before comparing these approaches, a terminology distinction is needed for describing final model structures rather than the chronological sequence of carbonisation.
We use \textit{cross-links} in the narrow chemical sense: aliphatic or heteroatom-bearing bridges connecting otherwise distinct aromatic domains. 
In polymer-like networks, such bridges can control swelling, chain or domain mobility, elastic response, and resistance to structural relaxation through their density and flexibility.
At the low H/C ratios typical of high-temperature biochars, however, these bridges become increasingly unlikely because the available hydrogen budget is small and aliphatic chains are thermally unstable.
The resulting low-H/C structures are therefore better described in terms of \textit{rigid aromatic interconnections}: fused-ring junctions, non-hexagonal rings, grain boundaries, dislocations, and disclinations that connect or distort aromatic domains within the carbon skeleton.
These interconnections may also influence swelling and stiffness, but through rigid network topology, stacking frustration, and resistance to graphitisation rather than through flexible bridge-like cross-link density.
This distinction is not only definitional. It is the molecular-scale mechanism behind the mechanical and biological resistance question raised later (Section \ref{sec5}, Question 2), and it should be read into the carbon-framework and interfacial/mechanical rows of Table \ref{tab:properties}, where ``network connectivity'' and ``fracture tendency'' become H/C-dependent in kind, not only in degree.

\textbf{Top-down models} are experimentally guided. 
The approach originates in coal science, where Mathews and co-workers constructed experimentally constrained atomistic models of coals, chars, and cokes.\cite{FernandezAlos2011} 
Their construction logic -- constraining elemental composition, functional-group abundances, and aromatic cluster sizes obtained from HR-TEM lattice-fringe analysis, then assembling fragments into a three-dimensional structure -- became the template for disordered-carbon models more broadly. 
A prominent application of the top-down strategy is the work of Ungerer \textit{et al.}, who used ultimate analysis, XPS, and $^{13}$C NMR to establish a suite of six representative kerogen macromolecular models capturing distinct organic matter types and thermal maturity levels.\cite{ungerer2015molecular}
In practice, a widening set of characterisation data -- solid-state $^{13}$C NMR, XPS, XRD, FT-ICR MS, gas adsorption, and HR-TEM -- bounds fragment selection without uniquely determining the assembled network.\cite{xiao2017direct,xiao2016h} 
Recent work has made this approach substantially more quantitative. 
Wood \textit{et al.} established a quantitative protocol for woody biochar model construction in which experimentally derived ensemble-average chemical descriptors are used to build the model, while independent physicochemical properties and HR-TEM morphology provide validation.\cite{Wood2024a,Wood2024b}
Ngambia \textit{et al.} extended this framework by introducing explicit control of microporosity through \textit{virtual voids}, a construction device that emulates the void space generated as volatile products are released during pyrolysis, and showed that this permits identification of the structural motifs responsible for pore assembly and reproduction of cumulative pore-size distributions in a top-down biochar model.\cite{Ngambia2024}
Related virtual-void concepts have also been used in bottom-up quenched MD to generate low-density microporous carbon structures,\cite{luo2021virtual} so the device is better viewed as a transferable construction constraint than as exclusive to one workflow.
Together, these studies show that, at least for well-characterised woody and cellulosic systems, experimentally constrained biochar models can be constructed reproducibly and tested against observables not used directly in model generation.

The strength of experimentally guided construction is its direct empirical anchoring, but that anchoring is only as reliable as the interpretation of the proxies it rests on.
Aromaticity derived from solid-state NMR, Raman, XPS, or elemental ratios depends heavily on deconvolution choices that propagate directly into model construction.
Even with precise spectral interpretation, macro-level constraints like density, porosity, and cluster size leave the exact connectivity of the aromatic network under-determined.
In top-down models, such connectivity is usually a consequence of the selected building blocks and assembly protocol, not a quantity specified or validated in its own right.
A second limitation is that construction and validation can become partly circular when the same derived descriptors, such as adsorption-derived pore-size distributions or deconvoluted XPS functional-group assignments, are used for both purposes.

\vspace{1.0em}

\textbf{Bottom-up models} are mimetic. 
With reactive force fields, MD and kinetic MC  simulations can mimic pyrolysis and its products.
They start from biomass precursors or chemically defined mixtures representing early pyrolysis intermediates, and allow fragmentation, dehydration, condensation, early cross-linking, volatile release, aromatisation, and network disorder to emerge under simulated thermal conditions.
By contrast, workflows that select product-like fragments and assemble them into structures consistent with experimental descriptors are better classified as top-down or hybrid approaches.
Bottom-up models, therefore, capture the mechanistic origins of structural features, including why certain pore geometries form, how heteroatoms incorporate, and what governs resistance to graphitisation.

The most complete demonstration of the bottom-up approach for a chemically analogous system remains Leyssale and co-workers' simulation of the geological conversion of cellulose, lignin, and algae into kerogen.\cite{Leyssale2017, Leyssale2022, Leyssale2023}  
Using reactive MD within an accelerated replica-exchange framework, with systems of 2000--7000 atoms across roughly 100 replicas, they tracked the full transformation -- from carbohydrate fragmentation through aromatisation -- capturing both the evolving carbon nanostructure and the associated gas phase. 
The parallel with biochar pyrolysis is direct: biomass precursors, thermal carbonisation, disordered aromatic carbon, and heteroatom incorporation. 
A recent review argues that biochar modelling should inherit this coal and kerogen toolkit rather than reinvent it.\cite{SierraJimenez2023review}

A critical missing ingredient has been a chemically realistic starting point, as prior simulations relied on idealised or single-component precursors rather than true biomass feedstocks.
The model of Addison \textit{et al.} is the most structurally complete molecular representation of lignocellulosic biomass currently available,\cite{addison2024atomistic} and a natural starting point for extending the reactive framework from analogue systems to a genuine biomass precursor (Figure \ref{fig:approaches}). 
The obstacle is scale: the full plant model contains $\sim$250,000 atoms, roughly two orders of magnitude larger than the kerogen systems, and maintaining experimentally meaningful thermodynamic conditions at that size remains an open challenge.

\vspace{1.0em}

\textbf{Hybrid models} fuse experimental constraints and mechanistic information to generate structure-resolved molecular ensembles that remain chemically plausible and experimentally grounded (Figure \ref{fig:approaches}).
In biochar modelling, this strategy integrates multimodal characterisation data with structural information derived from reactive MD and first-principles calculations.
Sierra-Jimenez \textit{et al.} built on Mathews' coal-construction framework introduced above, adapting it to generate atomistic representations of cellulose and woody biochars, replacing HR-TEM lattice-fringe constraints, which are less diagnostic for disordered biochar, with BPCA-derived aromatic cluster-size distributions.\cite{SierraJimenez2024}
Within this workflow, experimental measurements constrain key ensemble properties, including elemental composition, aromatic cluster-size distributions, and functional-group abundance, while reactive MD simulations of biomass pyrolysis provide mechanistic insight into plausible pyrolysis-derived fragments, location of functional groups, and inter-cluster linkages (cross-links). 
The resulting models were validated against EPR and LDI FT-ICR MS. 
In a complementary effort, Sierra-Jimenez \textit{et al.} constructed a first-principles and neural-network-driven spectral database for biochar building blocks, covering Raman, XPS, FTIR, and NMR signatures, \cite{sierra2025first} that enables fragment-level spectroscopy validation within the same model-construction framework. 
Although this hybrid approach provides a direct link between molecular structure, pyrolysis chemistry, and experimental characterisation, it remains computationally demanding because quantum-chemical predictions must be performed separately for each molecular fragment. 
Its main contribution is to show that such workflows can be made reproducible across independent research groups.

A well-established hybrid reconstruction paradigm is reverse MC and its hybrid variant, in which atoms are displaced to reproduce an experimental structure factor or PDF while an interatomic potential penalises energetically unreasonable configurations, suppressing the unphysical structures that unconstrained data-fitting can produce.\cite{jain2006molecular}
A complementary route builds structural models against a database of independently simulated three-dimensional porous carbons, then fits the experimental adsorption isotherm as a linear combination of their contributions. 
Vallejos-Burgos \textit{et al.} demonstrated this for nanoporous carbons, generating a kernel of 78 atomistic structures spanning a wide range of pore sizes and geometries, and recovering three-dimensional pore architecture from experimental isotherms without a one-dimensional slit-pore approximation.\cite{Vallejos2023} 
This isotherm-kernel method has not yet been applied to biochars, but it is directly relevant.

\vspace{1.0em}

\textbf{Limitations shared across approaches.} 
Three limitations cut across all three routes. 
The first is that most current models -- top-down, bottom-up, or hybrid -- represent only the organic carbon phase. Mineral particles embedded in or associated with the carbon matrix, and the carbon--mineral interface itself, are commonly omitted.
Bulk ash content and mineral elemental composition are routinely accessible, and crystalline phases can be identified by XRD, whereas the amorphous fraction and spatial carbon--mineral interface are harder to resolve, although synchrotron and spectromicroscopy methods increasingly constrain them. 
The gap is therefore not only analytical but representational: current atomistic models rarely translate these measurements into explicit mineral-inclusive structures. 
Consequently, models validated against gas adsorption isotherms or true density are often compared with properties that reflect both phases while accounting for only one.\cite{Wood2024b} 
For high-mineral-content biochars, this is not a minor simplification and should be stated explicitly. 
In principle, bottom-up workflows offer the clearest mechanistic framework for how carbon--mineral contacts arise during pyrolysis; in practice, the relevant phase transformations, diffusion, and interfacial reactions span timescales beyond accessible simulations and are not yet reliably described by available force fields.

The second limitation is scale. 
Mineral grains, inherited biomass pores, and particle-level transport pathways can be orders of magnitude larger than atomistic simulation boxes; representing them as undersized nanoscale inclusions would overemphasise surface contributions, introduce finite-size effects, and misrepresent the material hierarchy.\cite{hyvaluoma2018quantitative} 
As an order-of-magnitude guide, classical MD can typically access $10^5$--$10^6$ atoms -- simulation boxes up to tens of nanometres -- over tens to hundreds of nanoseconds, whereas reactive MD is more often limited to $10^3$--$10^5$ atoms over picosecond-to-nanosecond trajectories.

The third limitation is force field fidelity and sampling. 
Reactive methods such as ReaxFF\cite{van2001reaxff} describe bond-breaking and bond-forming events, and are therefore essential for pyrolysis, oxidation, radical chemistry, and parts of ageing, but they are more demanding than non-reactive classical MD\cite{de2016graphitization} and are usually restricted to smaller systems and shorter trajectories. 
Sampling compounds the cost: because bond rearrangements occur on timescales inaccessible to direct MD, reactive simulations commonly rely on elevated temperatures, often above 2000 K, to generate enough effective reactive collisions within the simulated time, which can bias the chemistry when conditions depart from realistic pyrolysis environments. 
Replica-exchange reactive MD mitigates this by exchanging configurations across a ladder of temperatures, but the choice of ensemble carries its own trade-off. 
At constant volume, the high-temperature replicas develop elevated pressure: for kerogen this is physically reasonable, since kerogen forms under geological confinement, but for biomass pyrolysis, which proceeds under comparatively low pressure, the same conditions are less transferable and may distort product distribution, carbonisation pathway, and final structure.
At constant pressure this artefact is avoided, but the hot replicas instead risk a first-order liquid--gas phase transition that degrades sampling. Either way, acceleration conditions must stay close to realistic pyrolysis to keep the product distribution physical.
Non-reactive classical force fields carry the mirror-image trade-off. OPLS, CHARMM, and TraPPE\cite{jorgensen1996development, huang2013charmm36, martin1998transferable} provide more robust dispersion and electrostatics than reactive potentials, and therefore suit equilibrium structure, hydration, adsorption, ion complexation, and transport in fixed frameworks, but cannot change bonding topology. 
No current force field framework convincingly spans both regimes across biochar's organic, heteroatom-rich, and mineral complexity.

\vspace{1.0em}

Taken together, these approaches illustrate a common principle: the construction route, validation target, and downstream simulation method should be selected according to the scientific question (Figure \ref{fig:approaches}). 
Top-down models may be sufficient for adsorption screening or equilibrium structure--property analysis, whereas oxidative degradation, carbon--mineral interface evolution, and ageing require more mechanistic or hybrid treatments, because they depend on reaction pathways, evolving interfaces, and changing chemical functionality. 
The most practical route, therefore, extends these hybrid strategies: mechanistic simulations identify reactive motifs, functional groups, and connectivity patterns that are then embedded into experimentally constrained models and tested against measurements withheld from construction.



\afterpage{%
\clearpage
\begin{landscape}

\begin{table}[!hp]
\centering
\fontsize{7.pt}{8.pt}\selectfont
\setlength{\tabcolsep}{1.2pt}
\renewcommand{\arraystretch}{0.84}
\setlength{\aboverulesep}{0.3ex}
\setlength{\belowrulesep}{0.3ex}
\setlength{\extrarowheight}{0.4pt}

\begin{tabular}{@{}L{2.0cm}L{4.80cm}L{4.80cm}L{4.4cm}L{4.6cm}@{}}

\toprule
\textbf{Target} &
\textbf{Representation} &
\textbf{Validation observables} &
\textbf{Status / main gap} &
\textbf{Representative modelling approaches} \\
\midrule

\textbf{Chemical functionality} &
Elemental composition; functional-group identity and abundance; heteroatom speciation; protonation; reactive-site accessibility; radical site density; surface charge and point-of-zero-charge behaviour &
Elemental analysis; XPS; solid-state $^{13}$C NMR; FTIR; FT-ICR MS; EPR; potentiometric and Boehm titration; zeta potential; pH$_{\mathrm{pzc}}$; pH-dependent adsorption &
Partially represented. Bulk composition is tractable; site-specific speciation, protonation, and accessibility remain weakly constrained &
DFT for local motifs and binding; ML/spectral databases for assignment; classical MD for fixed protonation states; reactive MD for transformations \\
\midrule

\textbf{Carbon framework and porosity} &
Aromatic cluster size; stacking, curvature, disorder and defects; three-dimensional scaffold architecture; network connectivity; micropore volume, pore-size distribution and pore connectivity; true density &
Gas adsorption isotherms; He pycnometry; HR-TEM; BPCA; PDF; XRD; WAXS; SAXS; SANS; Raman; SEM; X-ray micro-CT; Hg porosimetry &
Increasingly tractable when model size and microporosity are controlled explicitly; macrostructure and particle-scale transport remain out of scope &
Experiment-constrained reconstruction; classical MD relaxation; MD/GCMC adsorption; database/ML screening; coarse-grained or continuum models for particle-scale transport \\
\midrule

\textbf{Interfacial and mechanical properties} &
Hydration and wettability; ion/molecule binding; elastic response; fracture tendency; thermal transport; competitive adsorption in water; contaminant speciation &
Contact angle; adsorption and uptake curves; nanoindentation; Hg intrusion and He pycnometry/porosimetry; thermal conductivity; before/after mechanical tests; competitive isotherms; extraction recovery; regeneration cycles &
Limited. Classical MD/MC covers local wetting, adsorption, and small-strain response; particle-level fracture and wet--dry cycling require multiscale treatment &
Classical MD/MC for wetting, adsorption, and local mechanics; DFT for binding motifs; coarse-grained and continuum methods for fracture or particle-scale response \\
\midrule

\textbf{Reactive, redox \& transport behaviour} &
Oxidative and thermal recalcitrance; carbon persistence; radical chemistry; catalytic and redox activity; electron/proton transfer; adsorption and diffusion kinetics &
Proximate analysis/fixed carbon; atomic H/C$_{\text{org}}$ ratio; R50 thermal-recalcitrance index; TPO; TGA; EPR; electrochemistry; thermal and electronic resistivity; time-resolved uptake; catalytic activity/selectivity &
Poorly represented. Recalcitrance proxies are routinely measured but not yet predicted from structure; reactive and charge-aware methods remain limited by system size, sampling, and force field fidelity &
DFT, reactive MD, kinetic MC, and ML potentials for bond-changing or charge-aware processes; classical MD for diffusion in fixed frameworks \\
\midrule

\textbf{Mineral and ash components} &
Ash content and phase; crystalline/amorphous mineral matter; carbon--mineral interface geometry; nutrient and ion-exchange sites &
Ash content by combustion; XRD; SEM-EDX; TEM-EDX; XANES/synchrotron spectromicroscopy; ICP-OES/MS; $^{29}$Si and $^{31}$P NMR; cation exchange capacity; ion-release measurements; pH-dependent sorption &
Largely absent. Representative carbon--mineral interfaces and validated cross-interactions are major missing components for soil-relevant models &
DFT for interface motifs; classical MD with mineral force fields; hybrid carbon--mineral parameterisation; reactive or ML methods for dissolution/precipitation subproblems \\
\midrule

\textbf{Ageing and environmental state} &
Surface oxidation; carboxyl/hydroxyl accumulation; time-evolving oxidative reactivity and stability; mineral dissolution/re-precipitation; pore occlusion; organo-mineral association; biofilm/microbial interface &
Sequential XPS and FTIR; Boehm titration; cation exchange capacity evolution; time-resolved gas adsorption; TPO; accelerated artificial ageing; CO$_2$-mineralisation incubations; before/after microscopy &
Largely absent. Most models describe fresh, static biochar rather than an evolving sequence of aged material states &
Reactive MD for oxidation steps; sequential classical MD for hydrated/oxidised states; kinetic or coarse-grained models for long-timescale trajectories; hybrid workflows \\
\bottomrule

\end{tabular}

\caption{Scope map for molecular and atomistic biochar models. 
The required level of description depends on the scientific question; no single model is expected to capture all targets simultaneously. 
Quantities such as pore-size distribution and surface area are interpreted descriptors, so stronger validation compares simulated and measured curves, spectra, or images directly.
Failed observables should also be reported because they define the model's domain of applicability.}
\label{tab:properties}

\end{table}
\end{landscape}
\clearpage
}


\section{What does a model need to capture?}\label{sec4}

Table \ref{tab:properties} organises the properties a biochar model may need to reproduce into six linked targets: chemical functionality, carbon framework and porosity, interfacial and mechanical properties, reactive, redox and transport behaviour, mineral and ash components, and ageing and environmental state.
For each target, the table identifies the key features that must be represented, the observables that can validate them, and the main current modelling gap.

Across its rows, the table exposes the present imbalance: static carbon-framework properties are increasingly tractable, whereas the properties most relevant to soil performance and permanence -- reactivity, mineral interfaces, and ageing -- remain weakly represented.
The table should therefore be read as a scope map rather than a universal checklist: it identifies which claims require which observables, and where current modelling capability remains underdeveloped.

Within the organic carbon phase to which most current models remain limited (Section \ref{sec3}), these models perform reasonably well for structural properties when built at a sufficient scale.\cite{Ngambia2024,Wood2024b,SierraJimenez2024}
Chemical functionality is still under-represented. 
Bulk atomic H/C and O/C ratios are straightforward to impose, but site identity, functional-group density, protonation, spatial accessibility, and heteroatom speciation are much harder to validate. 
This is where model construction and force field development meet.
By contrast, mineral and ash components and ageing trajectories remain weakly represented, despite being central to soil amendment and carbon-permanence applications. 

Thus, which omissions matter depends on the application: for some adsorption-screening questions in low-ash woody chars, ignoring minerals may be acceptable; for soil amendment, nutrient release, or permanence, the omission is less defensible.
The same scale ceiling described in Section \ref{sec3} sets a further, feedstock-specific limit here: woody-derived biochars retain aspects of the parent biomass architecture -- cell-wall structure, vessel lumens, and macropores at tens to hundreds of micrometres -- that require particle-scale or continuum descriptions and cannot be captured by enlarging an atomistic box by a few nanometres.\cite{hyvaluoma2018quantitative}

Structural properties are tractable.
Early biochar models -- representing the material as a single aromatic disk or a small fragment -- produce numerical structural metrics, but the values are insensitive to the pore geometry that governs adsorption. 
Capturing microporosity requires two things: models large enough for pores to form between building blocks, and construction protocols that explicitly control pore volume and size distribution. 
As a working geometric heuristic, the simulation-box should exceed the widest represented pore by at least the non-bonded cut-off, so that a pore does not interact with its periodic image, i.e., for micropore-dominated biochars this implies box sizes of several nanometres and structures of order $10^4$ atoms or more.
Only recently have such models been demonstrated for biochar by independent groups,\cite{Ngambia2024, Wood2024b, SierraJimenez2024} and even then, the assumptions embedded in standard experimental pore-characterisation techniques -- slit-pore geometry, independent pore filling and idealised accessibility -- remain largely unchallenged at the biochar scale.\cite{Vallejos2023}

This is not a failure. It reflects where the field started: structural models were a necessary foundation.
The implication is that the next modelling advances must target those underdeveloped categories, rather than merely refining static carbon-framework representations.

Here the validation principle of Section \ref{sec2} can be made explicit as a hierarchy.
We distinguish five validation tiers of increasing stringency: 
\begin{description} 
\item[Tier 0 -- Composition:] elemental H/C$_{org}$, O/C$_{org}$, and ash content (necessary but weakly discriminating). 
\item[Tier 1 -- Bulk structure:] true density, XRD/PDF, and HR-TEM fringe statistics or BPCA aromatic distributions (discriminates coarsely). 
\item[Tier 2 -- Directly simulated observables:] simulated N$_2$, Ar, or CO$_2$ isotherms and simulated XPS, Raman, FTIR, or NMR spectra compared against their measured counterparts, rather than against descriptors derived from them. 
\item[Tier 3 -- Dynamic and kinetic:] uptake kinetics, multicomponent selectivity, diffusion coefficients, pH-dependent sorption, and cation-exchange capacity. 
\item[Tier 4 -- Evolution:] sequential oxidation or ageing trajectories compared against accelerated-ageing or incubation series. 
\end{description} 
Tiers 0--2 test static structure with increasing stringency; 
Tiers 3--4 test dynamic and time-evolving behaviour, and are therefore different in kind rather than simply more demanding. 
Higher tiers are thus not automatically ``better'' -- they answer different questions. 
The reporting rule is simple: state the tier at which a model has been tested, and do not present a Tier 1 model as validated for a Tier 3 question.

The pore-size distribution makes the Tier 1--Tier 2 boundary concrete: computing it from a model and comparing it to one inferred from a gas adsorption isotherm applies the same chain of assumptions twice. 
A stronger test computes the simulated adsorption isotherm directly and compares it to the measured curve -- more demanding, but also more informative.
That advantage is also chemistry-dependent.
For purely physisorbing probes such as N$_2$ or Ar, a direct isotherm test isolates pore geometry and accessibility.
When the gas can chemisorb at specific heteroatom sites -- CO$_2$ at basic nitrogen, or acid gases at oxygenated edges -- the simulated isotherm additionally probes surface chemistry, making it a more stringent test but also exposing a force field limitation: non-reactive potentials cannot reproduce site-specific chemisorption, so the model, not the method, sets the ceiling on what direct isotherm matching can validate.
The same caution applies to surface chemistry: XPS and Raman constrain bonding and disorder but do not uniquely measure aromaticity, while Boehm titration provides an operational acid-group inventory that complements the surface-sensitive, deconvolution-dependent picture from XPS.

This principle has been applied in practice: the iterative construction approaches of Wood \textit{et al.} and Ngambia \textit{et al.} validate against true density and HR-TEM images -- independent observables -- rather than BET surface area or derived pore-size distributions, both of which encode additional structural model assumptions. 
Vallejos-Burgos \textit{et al.} fitted experimental isotherms directly against a kernel of simulated three-dimensional carbon structures.\cite{Vallejos2023} 
A related but distinct standard is mechanistic complementarity between simulation and experiment.
For instance, Ngambia \textit{et al.} used atomistic models to identify manganese adsorption mechanisms, coordination environments, and functional-group preferences that cannot be resolved from batch and column experiments alone, while the experimental results provided the macroscopic performance context needed to interpret the simulations.\cite{Ngambia2026}


\section{Seven open questions for useful biochar models} \label{sec5}

The applications outlined earlier generate concrete modelling requirements.
The seven questions below follow directly from the imbalance summarised in Table \ref{tab:properties}: carbon-framework structure is now well described, whereas dynamic and reactive behaviour, mineral phases, and temporal evolution remain poorly represented. 

Each question, therefore, points to a distinct modelling and validation challenge.
These questions deliberately combine foundational uncertainties, such as persistence, ageing, and surface accessibility, with application-driven tests, such as catalysis and gas separation, because both expose the same missing molecular descriptors.
Table \ref{tab:crosswalk} maps these questions to the model objectives they stress (Table \ref{tab:properties}), the validation tier at which each must ultimately be tested (Section \ref{sec4}), and the community priorities that enable them (Section \ref{sec6}), so the roadmap reads as one connected structure rather than three parallel lists.

\begin{table}[!ht]
\centering

\setlength{\tabcolsep}{3pt}
\renewcommand{\arraystretch}{1.1}
\begin{tabular}{@{}L{3.0cm}L{3.4cm}L{1.8cm}L{2.0cm}@{}}
\toprule
\textbf{Open Question (Sec. \ref{sec5})} &
\textbf{Table \ref{tab:properties} targets} &
\textbf{Validation Tier (Sec. \ref{sec4})} &
\textbf{Priorities (Sec. \ref{sec6})} \\
\midrule

Q1. Carbon persistence &
Reactive/redox \& transport; ageing; mineral \& ash &
Tier 4 & I, III, IV \\
\midrule
Q2. Mechanical \& biological breakdown &
Interfacial \& mechanical; carbon framework &
Tier 3 & I, V \\
\midrule
Q3. Heteroatom-controlled adsorption \& catalysis selectivity&
Chemical functionality &
Tier 2--3 & I, II \\
\midrule
Q4. Carbon--mineral interfacial chemistry &
Mineral \& ash; interfacial \& mechanical &
Tier 2--3 & I, III \\
\midrule
Q5. Molecular-sieve behaviour &
Carbon framework \& porosity; interfacial &
Tier 3 & I, IV \\
\midrule
Q6. Ageing \& surface oxidation &
Ageing \& environmental state; chemical functionality &
Tier 4 & I, III \\
\midrule

Q7. Surface--bulk boundary &
Carbon framework; chemical functionality; interfacial &
Tier 2--3 & III, IV \\
\bottomrule
\end{tabular}

\caption{Crosswalk linking the seven open questions (Section \ref{sec5}) to the model targets they stress (Table \ref{tab:properties}), the validation tier required to genuinely answer them (Section \ref{sec4}), and the community priorities that enable them (Section \ref{sec6}). 
The mapping is indicative rather than exclusive; most questions touch several targets. 
Force-field development (Priority I) recurs across almost every question, reflecting its status as the shared bottleneck, whereas the highest-tier questions -- persistence and ageing -- are precisely those requiring Tier 4 evolution data that current models cannot yet supply.}

\label{tab:crosswalk}
\end{table}

\vspace{1.0em}

\textbf{1. How long does biochar persist in soil?} 
Carbon sequestration depends on the persistence of biochar carbon over policy-relevant timescales -- decades to centuries, not years.\cite{lehmann2015persistence} 
Here, \textit{persistence} refers to long-term carbon retention relevant to permanence accounting, whereas \textit{longevity} refers to the shorter operational lifetime over which a biochar retains a target function under defined environmental or technological conditions.
Persistence is not a single intrinsic number assigned by feedstock or pyrolysis temperature alone. 
It emerges from the coupling between oxidisable carbon-site density, oxygen and water accessibility, particle fragmentation, mineral access and protection. 
A useful model should therefore not aim to predict a century-scale residence time directly from molecular simulation. 
Its immediate role is to identify molecular and nanoscale descriptors that can parametrise larger-scale decay models: oxidisable site density, hydrophilicity, oxygen accessibility, pore connectivity, and the (de)stabilising role of carbon--mineral association. 
These molecular descriptors are the natural inputs to the empirical persistence metrics already used experimentally, such as atomic H/C$_{org}$ ratio, the R50 thermal-recalcitrance index, and CO$_2$-mineralisation rates from accelerated ageing or incubation, which underpin current carbon-permanence accounting.

These proxies are already policy variables: the European Biochar Certificate and the emerging EU Carbon Removal Certification Framework use molar H/C$_{org}$ thresholds -- for example $<$0.7 for certification and $<$0.4 for the most durable class -- to assign permanence, and the IPCC 2019 Refinement assigns 100-year permanence fractions on the same basis.\cite{ebc_guidelines, ipcc2019refinement, lehmann2021biochar} 

A model that predicts the structure $\rightarrow$ H/C$_{org}$ $\rightarrow$ decay-rate mapping therefore has direct accounting relevance. The natural downstream targets are two-pool (labile/stable) exponential decay models and mean-residence-time meta-analyses fitted to incubation data.\cite{fang2014biochar, azzi2024modelling} 

Leyssale and co-workers showed that accelerated reactive workflows can follow long-timescale carbon transformation in kerogen-like systems,\cite{Leyssale2017} but translating such approaches into predictive oxidation or mineralisation rates for biochar remains unresolved. 
The timescale gap between accessible simulation and policy-relevant persistence is not a detail to be refined away; it is the central obstacle this question must solve.

\vspace{1.0em}

\textbf{2. What controls physical fragmentation and biological accessibility?} 
Mechanical fragmentation and biological degradability are related, but they are distinct problems. 
For low-H/C chars, physical resistance may depend less on aliphatic cross-links than on the rigid aromatic interconnections defined in Section \ref{sec3} -- fused-ring junctions, non-hexagonal rings, dislocations, and disclinations -- that stitch small graphenic domains into a continuous but non-graphitic skeleton.
Abrasion, wet--dry cycling, freeze--thaw stress, and tillage can create fresh surfaces and open previously inaccessible pore space even where the carbon framework is chemically resistant.
Biological utilisation depends on a different set of variables: hydration, pH-dependent charge, nutrient co-location, redox activity, and the formation of organo-mineral coatings that change accessibility at the surface.
A physical constraint frames the biological problem before any chemistry is invoked.
Microorganisms are of order a micrometres and their extracellular enzymes of order a few nanometres, whereas the micropores ($<$2 nm) that dominate biochar surface area are smaller still. 
Most of the internal surface is therefore inaccessible to the agents that would mineralise it, and biological degradation must proceed inward from external surfaces, macropores, and freshly fractured faces. 
This couples Questions 1 and 2: fragmentation that creates new accessible surface, rather than intrinsic carbon chemistry alone, may set the effective persistence rate.
Useful models must therefore couple elastic or fracture response to wettability, surface accessibility, and interfacial chemistry, rather than treating stability as a single scalar property. 
Adjacent carbon-modelling studies already provide precedents, including nanoindentation simulations of soot and vitrinite models and coarse-grained compression of carbon networks.\cite{pascazio2020exploring, liu2022molecular} 
These give mechanical validation targets for disordered-carbon structures. 
Nanoindentation, cyclic wetting tests, and incubation studies with before-and-after characterisation provide the experimental anchors for this problem, but they have not yet been integrated into molecular biochar workflows.

\vspace{1.0em}

\textbf{3. How does heteroatom speciation affect adsorption and catalytic activity?} 
Nitrogen, oxygen, and sulphur alter local polarity, proton affinity, pore geometry, and reaction energetics, but bulk heteroatom content is a poor proxy for site-specific behaviour. 
The relevant variables are speciation, protonation state, spatial distribution, and accessibility: pyridinic, pyrrolic, graphitic, carbonyl, phenolic, and carboxylate environments do not contribute equivalently. 
For catalysis, these sites may also immobilise active metal centres or stabilise adsorbed reactants and intermediates, so adsorption and catalytic function should not be separated from surface chemistry.
Current models often treat heteroatoms as uniform decorations on an otherwise fixed carbon scaffold, missing the synthesis-dependent way in which they reshape both chemistry and texture. 
Classical MD and DFT have proven useful for selected non-reactive surface-chemistry questions, including ion complexation at explicit oxygenated sites.\cite{Ngambia2026, bachs2024understanding} 
Ultimately, DFT and experimentally constrained reconstruction are needed to determine which sites are present and relevant. 
Recent model-construction studies also show that synthesis conditions alter heteroatom retention and aromatic cluster distributions in the resulting char.\cite{SierraJimenez2025} 
A strong near-term benchmark would be a matched family of models with comparable pore architecture but systematically varied edge-species distributions, tested against pH-dependent adsorption or catalytic selectivity.

\vspace{1.0em}

\textbf{4. What role do mineral and ash components play at the molecular scale?}
Biochar's mineral phase is not inert.
Dissolved carbonates, silicates, and phosphates shift local pH, alkalinity, and ionic strength at hydrated biochar surfaces; iron oxides can introduce redox-active sites; while Ca$^{2+}$ and Mg$^{2+}$ can coordinate to oxygenated carbon edges, mineral oxygens, and selected nitrogen-containing donor sites, forming cation-bridged carbon--mineral or carbon--ion--organic complexes.\cite{sowers2018spatial, varga2025role}
For many soil and aqueous applications, the chemically relevant site is therefore neither pure carbon nor pure mineral, but the carbon--mineral boundary where wetting, ion binding, precipitation, and dissolution intersect.
Current atomistic models almost never represent this interface explicitly.
The goal is not to reproduce every ash component present experimentally, but to build representative interfaces for the dominant phases in a given feedstock class. This approach is most appropriate for materials in which the organic carbon matrix remains the dominant phase and ash or minerals act as embedded, surface-associated, or interfacial components. When the inorganic fraction becomes comparable to or exceeds the carbonaceous fraction, 
for example above roughly 50 wt.\% mineral matter, an organic-chemistry-centred construction workflow ceases to be the appropriate starting point, and a mineral-composite treatment becomes more suitable.
We stress that this is a \textit{modelling} boundary, not a materials-classification one: certification schemes such as the European Biochar Certificate deliberately removed a former minimum organic-carbon requirement to accommodate high-ash crop-residue and secondary-biomass chars,\cite{ebc_guidelines} 
so a material on the mineral-composite side of this modelling boundary remains biochar for certification purposes.
Force-matched or clay-style potentials for common phases such as quartz-like silicates, calcite, phosphates, and iron oxides provide a transferable starting point.\cite{Cygan2004,heinz2013thermodynamically} 
Mineral-inclusive kerogen and organic-shale models provide useful neighbouring examples, but the open challenge for biochar is assembling heterogeneous carbon--mineral interfaces 
at realistic composition and validating them against elemental mapping, ion release, and pH-dependent sorption behaviour.\cite{xie2026three} 

\vspace{1.0em}

\textbf{5. Can biochar's pore geometry achieve molecular sieve behaviour?}
Gas separation is a kinetics question as much as a thermodynamic one. 
Average pore geometry, therefore, indicates only a thermodynamic opportunity, not a molecular-sieve mechanism.
A pore may adsorb a gas strongly yet fail as a molecular sieve if the entrance aperture is too large, too flexible, or blocked by water under operating conditions.
What matters is not only average pore size, but accessibility from an external surface, pore connectivity, aperture distribution, diffusion barriers, and chemical heterogeneity at the pore mouth.
Most current models still report average structural descriptors more readily than site-resolved adsorption and transport.
The tools to do better already exist in neighbouring disordered-carbon communities, where adsorption and transport are treated in explicitly three-dimensional heterogeneous pore networks rather than as assemblies of idealised slit pores.\cite{Obliger2023, de2017structural}
The next step for biochar is to validate models against multicomponent adsorption and uptake kinetics, not only single-gas equilibrium isotherms.\cite{corrente2025slit}
Beyond carbon analogues, well-characterised polymer families with contrasting pore architectures -- cross-linked networks, polymers of intrinsic microporosity, and non-porous controls -- offer useful reference systems for benchmarking how pore geometry alone maps onto uptake, although parsing their surface from bulk contributions raises the same accessibility problem discussed in Question 7.

\vspace{1.0em}

\textbf{6. How should models represent biochar ageing?} 
Ageing is not a single state -- it is a trajectory. Most current models describe only its starting point. 
Freshly produced biochar and field-aged biochar are not the same material, and representing ageing as a small correction to a pristine structure is not sufficient.
Surface oxidation accumulates carboxyl and hydroxyl groups, mineral phases dissolve or reprecipitate, pores are progressively occluded by precipitates, sorbed organic matter or biofilms, and new organo-mineral associations form. 
A practical first-generation route is an ageing ladder of discrete model states -- for example fresh, partially oxidised, mineral-associated, and pore-occluded -- benchmarked against controlled artificial ageing or incubation experiments using sequential XPS, FTIR, adsorption, microscopy, and ion-release measurements. 
Resolving this problem may require a combination of reactive MD for oxidation trajectories, classical MD for successive hydrated carbon--mineral interface states, and coarse-grained treatments for longer-timescale pore evolution. 
The modelling target is not a single-aged structure, but a reproducible trajectory through chemically and environmentally defined states.

\vspace{1.0em}

\textbf{7. Where does the surface end and the bulk begin?} 
For highly porous, disordered materials, the surface--bulk distinction is not sharp. 
In practice, ``surface'' is an operational category that depends on the probe molecule, solvent environment, timescale, and protonation state.
An external functional group may be sterically irrelevant, while an internal pore wall may dominate adsorption, redox chemistry, or ion binding.
Binary surface/bulk assignments, therefore, misrepresent both accessibility and reactivity. 
Useful models must replace that binary with accessibility classes -- accessible, conditionally accessible, and inaccessible under defined conditions -- and test them by explicit simulations of water, ions, and molecular probes in the pore network. 
In aqueous systems, for example, accessibility also depends on pH, ionic strength, competing solutes, metal ions, and contaminant speciation. These conditions affect surface charge, ion-exchange behaviour and preferred adsorption sites, and should therefore be specified in the model definition.
What is surface depends on the question being asked. Until models make that explicit, surface chemistry will remain a question of structural resemblance rather than prediction.

\vspace{1.0em}


\section{A roadmap for the community towards useful biochar models}\label{sec6} 

These questions cannot be solved by another isolated model.
The main barriers are no longer structural-model construction alone -- they are missing reactive and carbon--mineral force fields, limited treatment of ageing, fragmented metadata and inconsistent validation practice. Five priorities follow directly. 
Across all five priorities, models should be developed in dialogue with experimentalists so that simulated descriptors correspond to measurable observables rather than convenient internal variables.

\vspace{1.0em}

\textbf{I. Force field development.}
Force fields face two distinct challenges that require different solutions.
For large classes of non-reactive structure--function questions -- surface adsorption, interfacial interactions, ion complexation, and small-strain mechanical response -- classical force fields can already be adequate when benchmarked carefully.\cite{Wood2024b, Ngambia2026, Wood2026}
Reactive force fields, by contrast, are genuinely needed for bond-breaking and bond-forming events: pyrolysis, surface oxidation, radical chemistry, and parts of the ageing trajectory. 
The realistic path forward is therefore multiscale: classical MD for equilibrium structural and interfacial properties, reactive methods applied selectively where bond events are mechanistically essential, and coarse-grained or continuum models for particle-scale transport, fragmentation, and field coupling. 

Machine-learned interatomic potentials, particularly equivariant frameworks such as MACE,\cite{batatia2025foundation} 
offer a promising middle ground for well-defined subproblems such as oxidised carbon edges and carbon--mineral interfaces, where targeted electronic-structure training data can be generated. 
They are not a universal solution, however: in biochar their usefulness will depend on training sets that span disordered aromatic carbon, heteroatom chemistries, hydration, protonation states, and relevant bond-breaking pathways. 
A further force field gap, distinct from the reactive-chemistry problem, is the absence of validated cross-terms for the carbon--mineral interface -- specifically, interactions between aromatic carbon surfaces and common biochar mineral phases such as silicates, carbonates, iron oxides, and phosphates. 
This gap is not a fundamental barrier: the force-matched and clay-style potentials noted for Question 4 (Section \ref{sec5}) already exist, and the clay--organic matter community has established the methodology for fitting carbon--mineral interactions from \textit{ab initio} data.
Adapting that methodology to biochar's specific mineral assemblages is a tractable near-term goal, with direct payoff for soil chemistry, adsorption, and permanence models.

\vspace{1.0em}

\textbf{II. An open model database.}
A shared database is not only desirable -- it is already beginning to emerge. 
Existing repositories from the Erastova group (\href{https://github.com/erastova-group}{github.com/erastova-group}), WSU/Penn State (\href{https://github.com/BiocharModeling/BiocharAtomisticModels}{github.com/BiocharModeling}), MoleCraftHUB (\href{https://github.com/MoleCraftHUB}{github.com/MoleCraftHUB}), and the Leyssale kerogen database (\href{https://gitlab.ism.u-bordeaux.fr/jleysall/practical-kerogen-models}{gitlab.ism.u-bordeaux.fr/jleysall}) already provide coordinates, building blocks, construction workflows, simulation inputs, and experimental constraints for selected systems.
These resources show that reproducible model construction is achievable. They are the beginning of a database, not the database itself.
Their long-term utility will also depend on explicit stewardship, including named maintainers, persistent identifiers, versioning policies, contribution routes, and sustained hosting beyond the lifetime of individual projects.

The priority is to move from isolated study-specific repositories to an open, versioned, curated, and metadata-rich community resource spanning diverse feedstocks, heteroatom-rich systems, and eventually mineral-inclusive and aged biochars. 
To be useful, the database must store not only coordinates but also construction scripts, force field and charge assignments, intended application domain, validation targets and the tier at which each was tested (Section \ref{sec4}), and known limitations.
It should also record failed observables, non-transferable conditions and proxy-derived constraints, because these define the model's useful domain as clearly as successful validation does.
It should tag models by ash content, dominant mineral phases, ageing state, and hydration/protonation assumptions, so that high-ash and low-ash biochars from different feedstocks are not conflated. 
That level of metadata will enable cross-study comparison, computational screening, and machine-learning property prediction.

Beyond the models themselves, a standard reference suite of experimentally characterised benchmark biochars -- with agreed validation observables and corresponding, openly published molecular models -- would give experimentalists and modellers shared targets for intercomparison, validation and method development.
The UK Biochar Research Center Standard Biochar Materials are a natural basis: they are publicly distributed, extensively characterised, and documented in the literature.\cite{masek2018standard} 
We propose adopting a small spanning set -- for example softwood, straw- and husk-derived chars produced at contrasting highest treatment temperatures -- with their models and validation data linked to \href{https://www.charchive.org/}{Charchive}.

\vspace{1.0em}

\textbf{III. Shared classification and metadata standards.}
Biochar is still inconsistently described across communities, and the problem becomes sharper as independent groups build models from different feedstocks, construction rules, and validation targets.\cite{Wood2024b, Ngambia2024, SierraJimenez2024}
What the field needs immediately is a minimal shared vocabulary linking model and experiment: feedstock class, highest treatment temperature, heating rate, residence time where available, atomic H/C$_{org}$ and O/C$_{org}$ ratios, porosity descriptors, ash content, and dominant mineral phases. 
This modelling metadata layer would complement, rather than replace, existing biochar production, certification, and reporting standards by adding the information needed for model--experiment interoperability.

As an example, a woody biochar at 600\textdegree C and a grass-derived biochar produced at the same temperature can share similar carbon nanostructure while differing by an order of magnitude in ash content and entirely in mineral phase composition. 
A classification that does not capture this conflates materials with fundamentally different behaviour in soil. 
Ageing or environmental state -- including dry or hydrated phase, pH, ionic strength, incubation history and oxidation state -- should also be encoded separately from production state because freshly produced and field-aged biochars are not equivalent materials. 
Whether this framework ultimately takes the form of a biochar class or a richer metadata ontology, without it, the shared database cannot function as a cumulative community resource.

\vspace{1.0em}

\textbf{IV. Ensemble validation and uncertainty reporting.} 
Because experimental characterisation data typically underdetermine a unique atomistic structure, the practical output of any construction route is an ensemble of plausible models rather than a single exemplar (Section \ref{sec3}). 
This ensemble character is repeatedly invoked as a conceptual safeguard but is not yet operationalised as a reporting requirement. 

Two practices follow directly. 
First, the structure-to-structure variance within an ensemble should be reported alongside the mean value for any target property -- pore-size distribution, aromatic cluster size, adsorption energetics -- so that a prediction can be distinguished from noise inherent to the construction protocol itself. 
Second, the field needs an application-specific 
minimum defensible ensemble size, because current studies vary widely in how many independent structures are built and compared, and rarely justify that choice against the sensitivity of the target property to construction stochasticity.
As an opening position for community discussion, we suggest at least five independent structures where the target is a structural descriptor, and at least ten where the target is adsorption energetics or a transport property.

Together these imply a minimal reporting set: the mean, the structure-to-structure standard deviation, the number of independent structures, the convergence criterion, and -- where feasible -- the spread evaluated at the validation tier tested (Section \ref{sec4}). 
Without it, an ensemble average can carry a false impression of precision, and disagreement between research groups' models cannot be distinguished from ordinary construction variance.
This priority should therefore be implemented jointly with Priority III: these fields should become required metadata in the shared database, not optional supplementary information.

\vspace{1.0em}

\textbf{V. Education and best practices.}
Molecular simulation tools are increasingly accessible. 
That is beneficial, but it also lowers the barrier to producing confident-looking wrong answers from poorly constrained models. 
The community should invest in modelling schools, mentorship networks, and shared protocols. 
These should be built around openly documented workflows and training resources, including current repositories such as the Erastova group tutorials (\href{https://github.com/Erastova-group/Biochar_Tutorials}{github.com/Erastova-group/Biochar\_Tutorials}) and the WSU/Penn State construction code (\href{https://github.com/BiocharModeling/BiocharAtomisticModels}{github.com/BiocharModeling/BiocharAtomisticModels}), as well as archived resources such as \href{https://carbonpotentials.org}{carbonpotentials.org}.\cite{de2019transferability}
\footnote{At the time of writing, \href{https://carbonpotentials.org}{carbonpotentials.org} is no longer live at its original domain, but remains accessible through the Internet Archive Wayback Machine: \href{https://web.archive.org/web/20240604135231/http://www.carbonpotentials.org/}{archived version, April 2024}. We flag this because the resource was a valuable reference for the community; its disappearance illustrates a broader sustainability problem for computational-science infrastructure, where useful workflows require long-term hosting, maintenance, versioning and funding to remain usable.}

Training should therefore teach not only how to run simulations, but how to choose a model for a question, identify unsupported assumptions, and recognise when apparent numerical precision exceeds experimental or force field reliability.
Training needs to become explicitly mineral-inclusive and ageing-aware: the next generation of mistakes will come not only from poor carbon models, but from treating biochar as a static, purely organic material when the application does not permit that simplification. 
Workshops such as the one behind this article (\href{https://www.cecam.org/workshop-details/advancing-biochar-molecular-models-from-experiments-to-model-construction-and-application-1375}{www.cecam.org}) 
provide a practical forum for sharing workflows, building common standards, fostering open discussion and identifying constructive steps forward.

Free, openly documented resources are essential, but they are not the only viable model for delivering training at scale.
Structured programmes such as the International Biochar Initiative's Virtual Graduate Lecture Series (\href{https://learn.biochar-international.org/p/virtual-graduate-lecture-series-biochar}{learn.biochar-international.org/virtual-graduate-lecture-series-biochar}) show what sustained, structured education looks like when the community invests collectively in it, and both models point to the same underlying requirement: quality training infrastructure depends on coordinated effort, not individual initiative.

\vspace{1.0em}

\begin{figure}
    \centering
    \includegraphics[width=1\linewidth]{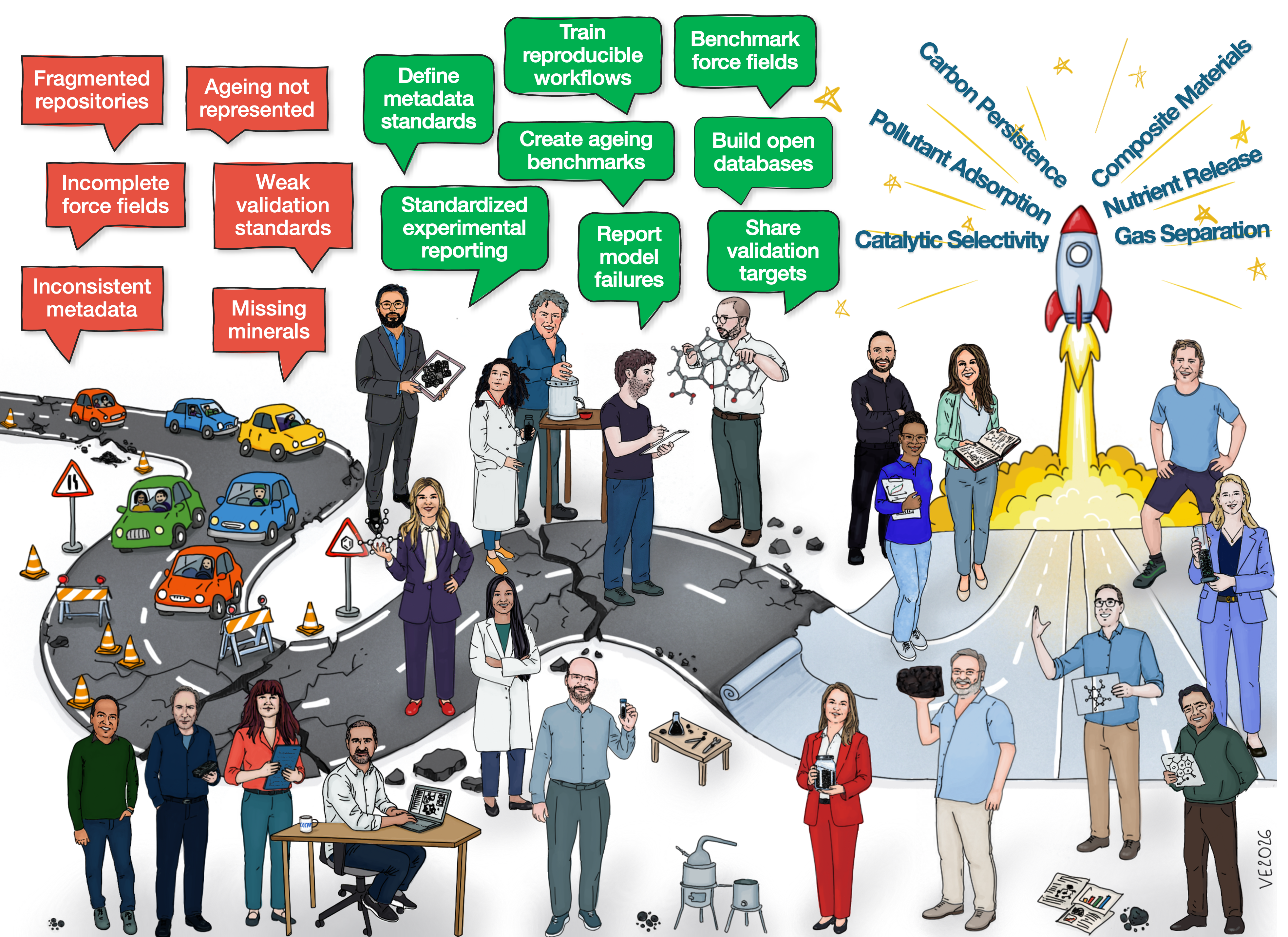}
    \caption{Community roadmap for advancing useful molecular models of biochar. 
    The hand-drawn figure illustrates some of the participants of the CECAM workshop, from which this article emerged, and summarises the collectively identified links between modelling bottlenecks, community actions and enabled target applications.}
    \label{fig:roadmap}
\end{figure}

Figure \ref{fig:roadmap} presents this roadmap as a community output: the hand-drawn format records the workshop origin of the synthesis, while the figure structure captures the modelling bottlenecks, infrastructure needs and application domains identified collectively.

\section{Conclusions and outlook}\label{conclusion}

The first generation of biochar molecular models answered the questions they were built to answer. That was necessary. 
The questions now driving the field -- carbon persistence and operational longevity, aqueous remediation, catalytic selectivity, nutrient release, gas separation, and ageing -- require models that go beyond a pristine organic carbon framework to incorporate mineral phases, carbon--mineral interfaces, environmental state and temporal evolution.

The barriers to that next generation are not conceptual. Force fields, validation strategies, construction workflows, and model repositories either already exist in neighbouring communities or are actively emerging in biochar. The biochar community does not need to repeat the trial-and-error of coal, kerogen, or activated carbon modelling. It needs to inherit it deliberately, extend it to the chemistries that matter for soil, and organise it into shared infrastructure.

A useful biochar model is built for a specific question, validated against independent observables, and explicit about the phase, environmental state, and timescale it represents. 
The field has the expertise. What it needs now are the shared standards and practice to make that expertise cumulative.

\section{Author contributions}

All authors contributed to the study conception and thematic design through participation in the CECAM Flagship Workshop.
The workshop was conceptualised, funded, and coordinated by VE, AN, JPM and MGP.
The discussions were led by JWM, JPM and VE, who conceptualised the initial framework and outline. 
Literature synthesis and drafting of specific sections were performed by VE, JPM, VSJ, JML, FJMM, CdT, LB and OM.
MGP, SG, RLJ, MMUH, MR, EB, DL, AN, KAH, FBEK, PG, AO, XZ, FO, CMG and JMT provided feedback on the revisions.
VE, VSJ and JPM integrated the sections and synthesised the final draft.
Figures and tables were prepared by VE and VSJ. 
VE hand-drew Figure 2, which features illustrated portraits of a few of the workshop participants (included with their consent) and developed the visual roadmap in consultation with the author group.
All authors participated in the critical revision of the manuscript for intellectual content, and all authors read and approved the final manuscript.

\section{Acknowledgements}

This work emerged from a flagship workshop supported and funded by CECAM. 
The authors thank CECAM and the local CECAM-HQ-EPFL team in Lausanne, Switzerland, for hosting and enabling the flagship workshop \href{https://www.cecam.org/workshop-details/advancing-biochar-molecular-models-from-experiments-to-model-construction-and-application-1375}{``Advancing biochar molecular models -- from experiments to model construction and application''}, held at CECAM-HQ-EPFL, Lausanne, Switzerland, on 4--6 June 2025.

We thank all workshop participants for the discussions that shaped the scope, priorities and community roadmap of this review:
Mohamed Hechmi Aissaoui, Davide Amato, Kutand Bayer, Luca Bellucci, Llu\'{i}s Blancafort, Edo Boek, Diego Camargo-Trillos, Andrea Dernbecher, Carla De Tomas, Anthony Dufour, Valentina Erastova, Marilyne Farhat, Isabel Fonts, Manuel Garcia-Perez, Stef Ghysels, Paola Giudicianni, Payam Ghorbannezhad, Corinna Maria Grottola, Kelly Hawboldt, Amir Jalalinejad, Apoorv Jain, Gabriela Duran Jimenez, Valentina Sierra Jimenez, Robert Johnson, Jean-Marc Leyssale, Wei Li, Diego Liberati, Stephanie MacQuarrie, Francisco Martin-Martinez, Jacob Martin, Ond\v{r}ej Ma\v{s}ek, Jonathan Mathews, James Kamau Mbugua, Nigel Marks, Audrey Lucrece Noumbissi Ngambia, Ama\"{e}l Obliger, Frederik Ossler, Muhammad Riaz, Juan Jesus Rico, Frederik Ronsse, Himanshu Sharma, Fredy Surahmanto, John Tobin, Mohammad Mezbah Ul Hoque, Lineya Vechnaya, Meenakshi Verma, Haryo Wibowo and Xiaolei Zhang.


\bibliography{bibliography}

\end{document}